\documentclass[aps,preprint]{revtex4}

\usepackage{amsmath,amssymb,amsfonts,graphicx,hyperref,amsthm}

\newcommand{\beq}{\begin{equation}}
\newcommand{\eeq}{\end{equation}}
\newcommand{\bea}{\begin{eqnarray}}
\newcommand{\eea}{\end{eqnarray}}

\begin{document}

\title{On the Global Casimir Effect in the Schwarzschild Spacetime}
\author{C. R. Muniz}
\email{celio.muniz@uece.br}\address{Grupo de F\'isica Te\'orica (GFT), Universidade Estadual do Cear\'a, Faculdade de Educa\c c\~ao, Ci\^encias e Letras de Iguatu, Iguatu, Cear\'a, Brazil.}

\author{M. O. Tahim}
\email{makarius.tahim@uece.br}\address{Grupo de F\'isica Te\'orica (GFT), Universidade Estadual do Cear\'a, Faculdade de Educa\c c\~ao, Ci\^encias e Letras do Sert\~ao Central, Quixad\'a, Cear\'a, Brazil.}

\author{M. S. Cunha }
\email{marcony.cunha@uece.br}\address{Grupo de F\'isica Te\'orica (GFT), Centro de Ci\^encias e Tecnologia, Universidade Estadual do Cear\'a, CEP 60714-903, Fortaleza, Cear\'a, Brazil.}

\author{H. S. Vieira}
\email{Horacio.Vieira@tufts.edu}
\address{Institute of Cosmology, Department of Physics and Astronomy, Tufts University, Medford, Massachusetts 02155, USA}
\address{Departamento de F\'{i}sica, Universidade Federal da Para\'{i}ba, Caixa Postal 5008, CEP 58051-970, Jo\~{a}o Pessoa, PB, Brazil}
\begin{abstract}

In this paper we study the vacuum quantum fluctuations of the stationary modes of an uncharged scalar field with mass $m$ around a Schwarzschild black hole with mass $M$, at zero and non-zero temperatures. The procedure consists of calculating the energy eigenvalues starting from the exact solutions found for the dynamics of the scalar field, considering a frequency cutoff in which the particle is not absorbed by the black hole. From this result, we obtain the exterior contributions for the vacuum energy associated to the stationary states of the scalar field, by considering the half-summing of the levels of energy and taking into account the respective degeneracies, in order to better capture the nontrivial topology of the black hole spacetime. Then we use the Riemann's zeta function to regularize the vacuum energy thus found. Such a regularized quantity is the Casimir energy, whose analytic computation we show to yield a convergent series. The Casimir energy obtained does not take into account any boundaries artificially imposed on the system, just the nontrivial spacetime topology associated to the source and its singularity. We suggest that this latter manifests itself through the vacuum tension calculated on the event horizon. We also investigate the problem by considering the thermal corrections via Helmholtz free energy calculation, computing the Casimir internal energy, the corresponding tension on the event horizon, the Casimir entropy, and the thermal capacity of the regularized quantum vacuum, analyzing their behavior at low and high temperatures, pointing out the thermodynamic instability of the system in the considered regime, {\it i.e.} $mM\ll 1$.\\

\vspace{0.75cm}
\noindent{Key words: Schwarzschild metric, Casimir effect, Singularity.}
\end{abstract}

\maketitle

\section{Introduction}

One of the physical phenomena which is an unequivocal manifestation of the vacuum associated with quantum fields is the so-called Casimir effect \cite{01}. In its original form, this effect occurs due to stochastic fluctuations in the vacuum expected value of the electromagnetic field and, thus, a finite vacuum energy arises because the presence of metallic boundaries, causing an attraction between them, despite the fact that they are electrically uncharged. As stated above, the effect occurs with any quantum field and is quite depending on the geometry under analysis \cite{Brown,Balian1,Balian2,Cognola,Elizalde1}. From a theoretical perspective, this phenomenon offers a large realm of studies for applied Mathematics \cite{Bordag1,Zhug,Kleinert1}. From the experimental point of view, measurements of the Casimir force have been performed with increasing precision \cite{Moste1,Rodriguez,Moste2}.

It was also remarked that the Casimir effect can occur as a consequence of a nontrivial topology of the physical space \cite{Moste3,Milton1,Bordag3,Moste4} or even of more abstract spaces, as those ones associated to quantum states \cite{Adolfo,Rodriguez2,Alberto}. In certain sense, the Casimir effect reflects a way to measure different topologies of a given system. Regarding the first kind of topological (or global) Casimir effect, some of these systems, from condensed matter to astrophysics and cosmology, have been extensively investigated \cite{Moraes1,Saharian,Eugenio,14,15,16,Celio1}. However, with respect to black holes, there are no studies related to the Casimir effect due to the spacetime topology itself, but just those ones which explore features of the spacetime geometry, with the imposition of external boundaries to the field \cite{Setare,Setare2,Sorge,Muniz2}. In addition, there exist the known difficulties to one define global energy in the context of general relativity, hence the use of local quantities (densities) in these works, as vacuum expectation values, in order to describe that effect.

The nontrivial topology of the black hole spacetime is essentially dictated by the presence of the singularity in it. Thus, we raise here the question: What is a singularity and how can we detect it? In particular, do black holes in fact ``pierce'' the spacetime? From a mathematical viewpoint this question was already addressed (a nice discussion of how to treat spacetime singularities can be found in \cite{Wald:1984rg}). The best way is just to analyse the holes left by the singularity and study the behavior of geodesics in this new manifold: Basically we use the singularities theorems. However the idea of adding a hole to the spacetime changes drastically its topology. From a physical viewpoint and in the context of semiclassical gravity there is a way to detect such a change of topologies, which is through the computation of the Casimir energy as we point in this work.

In this paper, we will study the quantum vacuum energy of an uncharged massive scalar field, initially at zero temperature, in a topologically nontrivial space, namely, in the background of the Schwarzschild metric. In order to do this, we will calculate the energy eigenvalues $\omega_n$ in the low frequency regime after finding the exact solutions for the field dynamics. These energies correspond to the stationary states of the field around the black hole. We will compute the half-sum of the eigenvalues of energy supposing that it captures better than certain local quantities features of the spacetime topology under consideration. Thus the called global Casimir energy at zero temperature associated with quantum fluctuations of the massless scalar field will be obtained through the expression $E_0=\frac{\hbar}{2}\sum_ng_n\omega_n$, where $n$ indexes the modes of the field and $g_n$ computes the degeneracies per mode. The vacuum energy calculated from this formula is divergent and then it will be regularized  by using the Riemann's zeta function procedure. Such a regularized quantity is the Casimir energy. It is worth point out that we will consider only the contribution due to the stationary states of the field for such a quantity, {\it i.e.}, those states which do not tunnel into the black hole. As it is known, the black hole singularity cannot classically manifest itself for anyone out of the event horizon, according to the cosmic censure hypothesis. We will show here that at least from a semi-classical point of view this is possible, and we advocate that the global Casimir effect permits, in principle, to identify the singularity.

Then, we will show that the Casimir tension on the event horizon due to that energy suggests the presence of the black hole singularity because a remaining value arises for such a tension when its mass vanishes. The same problem will be analyzed also at finite temperature through the computation of the Helmholtz free energy, from which we will obtain the Casimir internal energy, the corresponding tension on the horizon, and the Casimir entropy. Here a remaining tension arises again pointing to the presence of the singularity. We will also show that both Casimir tension and entropy reach constant and minimum values in the high temperature regime.

The present paper is organized as follows: In section II, we present the exact solutions for the scalar field around the black hole, calculate the Casimir energy as well as the tension on the event hoorizon in the low frequency limit, and analyze the thermal effects on the Casimir energy. Finally, in section IV, we conclude and close the paper.

\section{THE GLOBAL CASIMIR ENERGY}

Prior to study the global Casimir effect around a static black hole, we must briefly investigate the behavior of a massive scalar field in the background of the Schwarzschild solution.

\subsection{The field solution}

The solutions of a scalar field around the static black hole was already discussed in 1998 in the Frolov {\it et al.} book \cite{Frolov:1998wf}. In this book there are some references from the fifties about the earlier works on that subject. It is important to remark that those works considered only approximate solutions to the problem. The reason for these comments is the absence in the literature, until recently, of the exact and complete solutions for the field dynamics around these objects, a problem that has just been solved in \cite{Horacio1,Horacio2,CelioEPL} for the scalar field due to the development of the Heun functions studies in recent years.

In this way, we must solve the covariant Klein-Gordon equation, which is the equation that describes the behavior of scalar particles in the spacetime under consideration. In a curved spacetime, we can write the Klein-Gordon equation of a uncharged massive scalar particle coupled minimally with the gravity as
\begin{equation}
\left[\frac{1}{\sqrt{-g}}\partial_{\mu}\left(g^{\mu\nu}\sqrt{g}\partial_{\nu}\right)+m^{2}\right]\Psi=0\ ,
\label{eq:Klein-Gordon_cova}
\end{equation}
where we adopted the natural units $c \equiv \hbar \equiv 1$ and $m$ is the particle mass. On the other hand, the background generated by a static and uncharged black hole is represented by the Schwarzschild metric \cite{MTW:1973}, which in the Boyer-Lindquist coordinates \cite{JMathPhys.8.265} can be written as
\begin{equation}
ds^{2}=\left(1-\frac{2M}{r}\right)dt^{2}-\left(1-\frac{2M}{r}\right)^{-1}dr^{2}-r^{2}d\Omega^2,
\label{eq:metrica_Kerr-Newman}
\end{equation}
where $d\Omega^2=d\theta^{2}-\sin^{2}\theta\ d\phi^{2}$ and $M$ is the mass of the source, with $G \equiv 1$. In order to solve the Eq.~(\ref{eq:Klein-Gordon_cova}), we assume that its solution can be separated as follows
\begin{equation}
\Psi=\Psi(\mathbf{r},t)=R(r)Y_{l}^{m_l}(\theta,\phi)\mbox{e}^{-i\omega t}\ ,
\label{eq:separacao_variaveis}
\end{equation}
where $Y_{l}^{m_l}(\theta,\phi)$ are the spherical harmonic functions. Plugging Eq. (\ref{eq:separacao_variaveis}) and the metric given in Eq.~(\ref{eq:metrica_Kerr-Newman}) into (\ref{eq:Klein-Gordon_cova}), we obtain the following radial equation
\begin{equation}
\frac{d}{dr}\left[r(r-2M)\frac{dR}{dr}\right]+\left(\frac{r^{3}\omega^{2}}{r-2M}-m^{2}r^{2}-\lambda_{lm_l}\right)R=0,
\label{eq:mov_radial_1}
\end{equation}
where $\lambda_{lm_l}=l(l+1)$.

This equation has singularities at $r=(0,2M)$, and at $r=\infty$, and can be solved in terms of Heun-type equation, which solutions are
\begin{eqnarray}
R(x) & = &C_{1} \mbox{e}^{\frac{1}{2}\alpha x}x^{\frac{1}{2}\beta} \mbox{HeunC}(\alpha,\beta,\gamma,\delta,\eta;x) + C_{2}\mbox{e}^{\frac{1}{2}\alpha x}\ x^{-\frac{1}{2}\beta}\ \mbox{HeunC}(\alpha,-\beta,\gamma,\delta,\eta;x)
\label{eq:solucao_geral_radial_Kerr-Newman_gauge}
\end{eqnarray}
where $x=r-2M$ and $r$ runs over the range $2M <r <\infty $, $C_{1}$ and $C_{2}$ are constants, and the parameters $\alpha$, $\beta$, $\gamma$, $\delta$, and $\eta$ are given by \cite{Horacio2}
\begin{subequations}
\bea
\alpha&=&-4M\left(m^{2}-\omega^{2}\right)^{1/2}\\
\label{eq:alpha_radial_HeunC_Kerr-Newman}
\beta&=&i4M\omega\\
\label{eq:beta_radial_HeunC_Kerr-Newman}
\gamma&=&0\\
\label{eq:gamma_radial_HeunC_Kerr-Newman}
\delta&=&4M^{2}\left(m^{2}-2\omega^{2}\right)\\
\label{eq:delta_radial_HeunC_Kerr-Newman}
\eta&=&-l(l+1)-4M^{2}\left(m^{2}-2\omega^{2}\right).
\label{eq:eta_radial_HeunC_Kerr-Newman}
\eea
\end{subequations}

These two functions are linearly independent solutions of the confluent Heun differential equation provided $\beta$ is not an integer \cite{Ronveaux}.

In order to compute the global Casimir energy, we must calculate firstly the energies associated to the stationary states of the massive scalar field around the black hole. In order to do this, we need evaluate the natural boundary conditions on the field solutions at the asymptotic region (infinity), which in this case requires the necessary condition for a polynomial form of $R(x)$, since that the confluent Heun solutions have irregular singularities there. Following \cite{Fiziev}, we must impose the so called $\delta_N$ and $\Delta_{N+1}$ conditions, respectively
\begin{eqnarray}
\frac{\delta}{\alpha}+\frac{\beta+\gamma}{2}+1&=&-n \label{eq:cond_polin_1}\\
\Delta_{N+1}&=&0
\label{eq:cond_polin_2}
\end{eqnarray}
with $n$ a positive integer. Provided the fulfilment of the above two $\delta-$conditions, the confluent Heun solutions reduces to a polynomial of degree $N$,
as described there in Ref. \cite{Fiziev}.

The first condition Eq. (\ref{eq:cond_polin_1}) gives the following expression for the energy levels
\begin{equation}
n+1+i2M\omega-\frac{M\left(2\omega^{2}-m^{2}\right)}{\sqrt{m^{2}-\omega^{2}}}=0.
\label{eq:energy_levels}
\end{equation}
Now, we consider the low frequency limit, $\omega M\rightarrow 0$, which means that the particle is not absorbed by the black hole. In fact, the relative absorption probability of the scalar wave at the event horizon surface of a static and uncharged black hole is given by \cite{Horacio1}
\begin{equation}
\Gamma_{ab}=1-e^{-8\pi M\omega},
\end{equation}
and, in the considered limit, that probability approximates to zero. In this regime, the particle does not penetrate via tunneling the effective potential barrier that exists around the black hole (the Regge-Wheeler potential), in such a way that the reflection coefficient tends to the unity \cite{Frolov}. In fact, any stationary solution, namely, that with a real frequency is formed by waves that propagate outward from the event horizon superposing with the waves that tunnel out through the Regge-Wheeler barrier and move toward the horizon. Imposing the condition of no waves coming out from the horizon introduces complex valued frequencies. When $\omega$ is complex the solution corresponds to quasi-bound states.

The considered limit $M\omega \rightarrow 0$ also can be thought as a cutoff introduced in order to eliminate high frequencies, as usual in the analysis of the Casimir effect. It is worth yet notice that, for $m=0$ and in that approximation, we have the solution $n=-1$, which does not make any sense. This seems to suggest that, while classically orbits (unstable) for such a kind of particle exist just at $r=3M$ \cite{Frolov}, quantum mechanically they do not exist in any way, {\it i.e.}, there is no massless stationary states exterior to the black hole.

Then, we take into account the possible states which are stationary around the black hole and therefore the real eigenvalues of energy have the form
\begin{eqnarray}
\label{03}
E_{n}=m c^2\sqrt{1-\frac{G^{2}m^{2}M^{2}}{\hbar^2c^2n^2}},
\end{eqnarray}
for $n=1,2,3...$. It is worth point out that Eq. (\ref{03}) was also found in \cite{Barranco} by following a different approach. Notice that the bound energy $E_{n,b}=E_n-mc^2$ reaches a minimum equal to $-mc^2$ when $GM/c^2=n\hbar/mc$ or $R_S/2=n\lambda_C$, where $R_S$ is the Schwarzschild radius and $\lambda_C$ is the Compton length of the particle. Thus, in order to consider all the quantum states of the particle, {\it i.e.}, without a bounded $n$, we must have $\lambda_C\geq R_S/2$, which is compatible with the low frequency limit that we are using in here, implying real energies and therefore without the possibility of the tunnel effect to the black hole interior.

We have reintroduced the fundamental constants in Eq. (\ref{03}) in order to call attention to the fact that its non-relativistic approximation, $\mathcal{O}(1/c^2)\rightarrow 0$, after subtracting the rest energy of the particle, is exactly equal to the Bohr's energy levels of a ``gravitational hydrogen-like atom'', namely
\begin{eqnarray}
\label{03.1}
E_{n}\approx-\frac{G^2m^3M^2}{2\hbar^2n^2},
\end{eqnarray}
which also was obtained in \cite{Laptev,Doran}. This last result points out the consistence of our analysis until here.

\subsection{Regularized vacuum energy at zero temperature}

By considering Eq. (\ref{03}), we have that the quantum vacuum energy of the massive scalar field at zero temperature is
\begin{eqnarray} E^{(0)}=\frac{1}{2}\sum_{n=1}^{\infty}n^2\omega_n=\frac{m}{2}\sum_{n=0}^{\infty}n^2\sqrt{1-\frac{m^2M^2}{n^2}},
\label{04}
\end{eqnarray}
in which we again used the natural units. The factor $n^2$ comes from the degeneracy of the system. It is possible to show easily that the sum in Eq. (\ref{04}) is divergent, and in order to make that quantity finite, or regularized, we need adopt a regularization procedure here, and we choose to do this by means of the Riemann's zeta function. Firstly, we will use the binomial expansion of the square root so that
\begin{equation}\label{7}
E^{(0)}=\frac{m}{2}\sum_{n=1}^{\infty}n^2\left[1+\sum_{k=1}^{\infty}\binom{1/2}{k}(-1)^k\left(\frac{mM}{n}\right)^{2k}\right].
\end{equation}
Regularizing via the zeta function, we arrive at
\begin{equation}
E_{reg}^{(0)}=\frac{m\sqrt{\pi}}{4}\sum_{k=1}^{\infty}\frac{(-1)^k(mM)^{2k}\zeta(2k-2)}{\Gamma(k+1)\Gamma(3/2-k)},
\end{equation}
where we have used $\zeta(-2)=0$. This expression converges for $mM<1$. The series term corresponding to $k=1$ is the contribution  of the gravitational Bohr levels, as per Eq. (25), to the regularized quantum vacuum energy, which are non-relativistic and prevalent, therefore, in regions quite far from the horizon. Notice that we are not taking into account any scattered states which can occur in those regions.

We depict in Fig. 1 the Casimir energy per mass unity of the scalar particle as a function of the product $mM$ in units of Planck energy $(E_P)$. The maximum value is $E_{reg}^{(0)}\approx 0.015 E_P$ reached at $mM\approx 0.67$.

\begin{figure}[!h]
\centering
\includegraphics[scale=0.7]{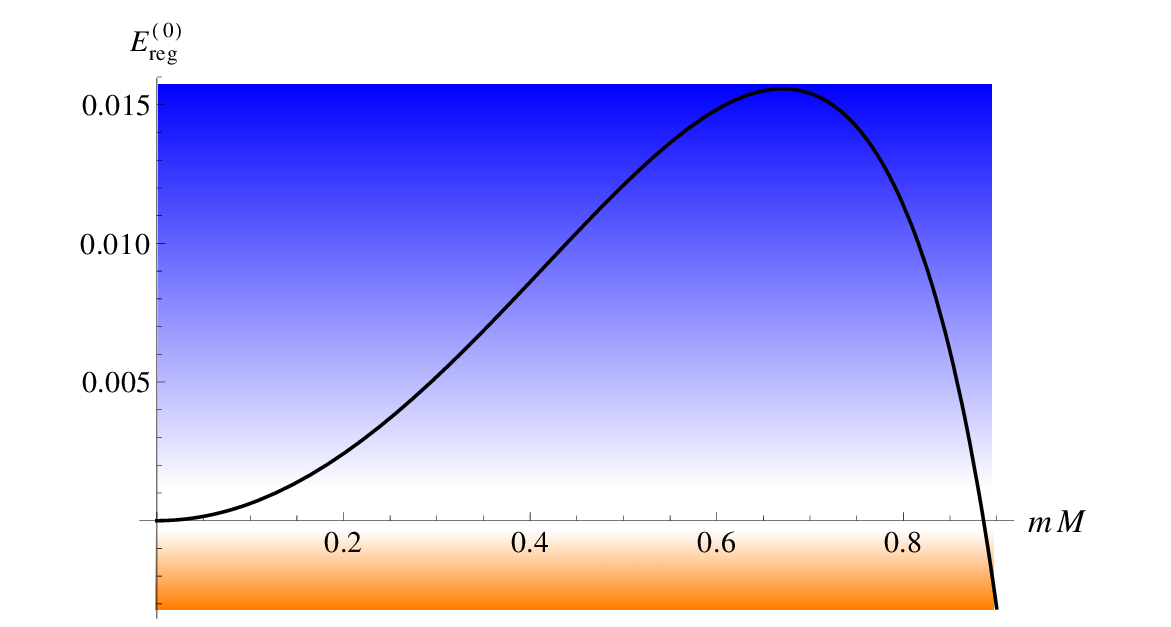}
\caption{Casimir energy per mass unity of the scalar particle as a function of the product $mM$, at $T=0$.}
\end{figure}

It is worth to point out that, in general, the Casimir energy depends on some geometrical parameter of the system. In our case, it is a function of the horizon radius, since the dependence on the black hole mass is in fact on the Schwarzschild radius, $R_h=2M$. In addition, that energy must cause a surface tension, which can be defined as the reversible work of formation of a unit area of surface and is given by $\tau=\partial E/\partial S$, where $S$ is the surface area. Thus, the tension on the horizon surface, of area $S_h=16\pi M^2$, is given by
\begin{equation}
\tau_h=\frac{\partial E_{reg}^{(0)}}{\partial S_h}=\frac{m}{64\sqrt{\pi}M^2}\sum_{k=1}^{\infty}\frac{(-1)^{k}k(mM)^{2k}\zeta(2k-2)}{\Gamma(k+1)\Gamma(3/2-k)}.
\end{equation}
We can see that the first term of the above series does not depend on the mass of the black hole, so that for $M\rightarrow 0$ this tension tends to $\tau_h=m^3/128\pi$. We consider this term as one that arises from the black hole singularity. Whereas the Casimir energy does not furnish any indication about the nontrivial topology of the spacetime given by the presence of the black hole singularity, the fact that the horizon surface tension is non-null when its mass goes to zero suggests the existence of that singularity, and therefore of the nontrivial topology of the spacetime under analysis. It is worth also notice that other nontrivial topologies without a singularity, as the surface of an ordinary sphere, does not present residual Casimir tension when its radius tends to infinity, {\it i.e.}, when the spacetime is ``flatted'' \cite{Moste3}.

\subsection{Temperature effects}

We analyze now the thermal corrections to the Casimir energy from the calculation of the Helmholtz free energy, given by \cite{Moste3}
\begin{equation} \label{FreeEnergy}
F^{(0)}=\beta^{-1}\sum_{n=1}^{\infty}n^2 \log{\left[1-\exp{\left(-\beta  \omega_n\right)}\right]}=\beta^{-1}\sum_{n=1}^{\infty}n^2 \log{\left[1-\exp{\left(-\beta m \sqrt{1-\frac{m^2M^2}{n^2}}\right)}\right]},
\end{equation}
where $\beta=1/k_BT$. The purpose here is verifying the behavior of these corrections when the mass of the black hole vanishes in order to see how the singularity appears. Therefore, we will consider the regime in which $mM\ll 1$ so that the free energy becomes
\begin{equation}\label{FreeEnergyApproxim}
F^{(0)}\approx -\beta^{-1}\sum_{n=1}^{\infty} n^2\sum_{k=1}^{\infty} \frac{e^{-\beta k m}}{k}\left(1+\frac{\beta m^3M^2k}{2n^2}\right),
\end{equation}
where we made the series expansion of the logarithm. Regularizing this quantity via Riemann's zeta function and considering again that $\zeta(-2)=0$ we get
\begin{equation}\label{FinalFreeEnergy}
F^{(0)}\approx- \sum_{k=1}^{\infty}e^{-\beta k m}\frac{m^3M^2}{2}\zeta(0)=\frac{m^3M^2}{4(e^{\beta m}-1)}.
\end{equation}
The thermal corrections to the Casimir energy - the internal energy - are given by
\begin{equation}\label{ThermalCasimirEnergy}
U^{(0)}(T)=-T^2\frac{\partial{F^{(0)}/T}}{\partial T}\approx \frac{m^3M^2}{4(e^{\beta m}-1)}-\frac{\beta m^4M^2}{16\sinh^2{(\beta m/2)}}.
\end{equation}
It is immediate to verify that these corrections go to zero when $M\rightarrow 0$. Here there are also no clues about the black hole singularity. However, if we take again into account the tension on the event horizon due to the Casimir thermal energy (\ref{ThermalCasimirEnergy}), in that limit we have the exact expression for the remaining tension
\begin{equation}\label{ThermalCasimirTension}
\tau_h(T)=\frac{m^3}{64\pi(e^{\beta m}-1)}-\frac{\beta m^4}{256\pi\sinh^2{(\beta m/2)}},
\end{equation}
which denotes the presence of the singularity. The high temperatures limit for this tension goes to a constant and minimum value given by $\tau_h(T)\approx-m^3/128\pi$, which is the opposite value to the case at zero temperature. Thus, in this regime the total tension on the horizon vanishes.
In the low temperatures limit the remaining tension is $\tau_h(T)\approx (m^3-\beta m^4) e^{-\beta m}/64\pi$. In Fig.2 we depict the remaining Casimir tension on the event horizon, as a function of the temperature, in Planck unities. Notice the curve evolution to a constant value at high temperatures.

\begin{figure}[!h]
\centering
\includegraphics[scale=0.7]{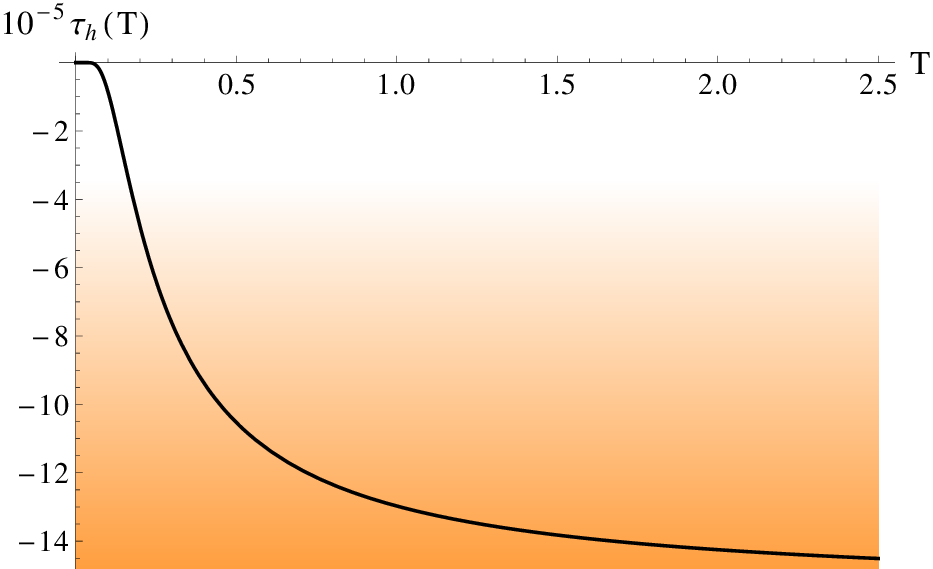}
\caption{Remaining thermal Casimir tension on the event horizon as a function of the temperature, in Planck unities, for $m=0.5 E_P$.}
\end{figure}

Another interesting quantity that we can calculate in the regime under consideration ($mM\ll 1$) is the Casimir entropy, $S^{(0)}=-\partial F^{(0)}/\partial T$, whose leading term is
\begin{equation}\label{entropy}
S^{(0)}\approx-\frac{k_B\beta^2 m^4M^2}{16\sinh^2{(\beta m/2)}}.
\end{equation}
In the low temperatures limit, the Casimir entropy is given by $S^{(0)}\approx -(1/16)\beta^2m^2M^2\exp{(-\beta m)}$, so that when $T\rightarrow 0$, $S^{(0)}\rightarrow 0$, obeying the third law of thermodynamics, therefore. In the high temperatures limit, the Casimir entropy tends to the constant value
\begin{equation}
S^{(0)}\approx -\frac{k_B m^2M^2}{4}.
\end{equation}
It is interesting to point out also that the Casimir entropy is proportional to the horizon area, exactly as in the Hawking entropy. The negative values found for the entropy indicate that we are not taking into account all the history. In fact, we are considering here only the entropy of the vacuum associated to the stationary states of the quantum field. On the other hand, the heat capacity at constant volume, given by
\begin{equation}
C_V=T\left(\frac{\partial{S^{(0)}}}{\partial T}\right)_V\approx\frac{m^4 M^2 \text{csch}^2\left(\frac{m}{2 k_B T}\right)}{8 k_B T^2}-\frac{m^5 M^2 \coth \left(\frac{m}{2 k_B T}\right) \text{csch}^2\left(\frac{m}{2 k_B T}\right)}{16 k_B^2 T^3},
\end{equation}
is always negative, according to the FIG.3 below, which points to the thermodynamic instability of the quantum vacuum under consideration, besides the positive values for the Helmholtz free energy given in Eq. (\ref{FinalFreeEnergy}).

\begin{figure}[th]
\centering
\includegraphics[scale=0.7]{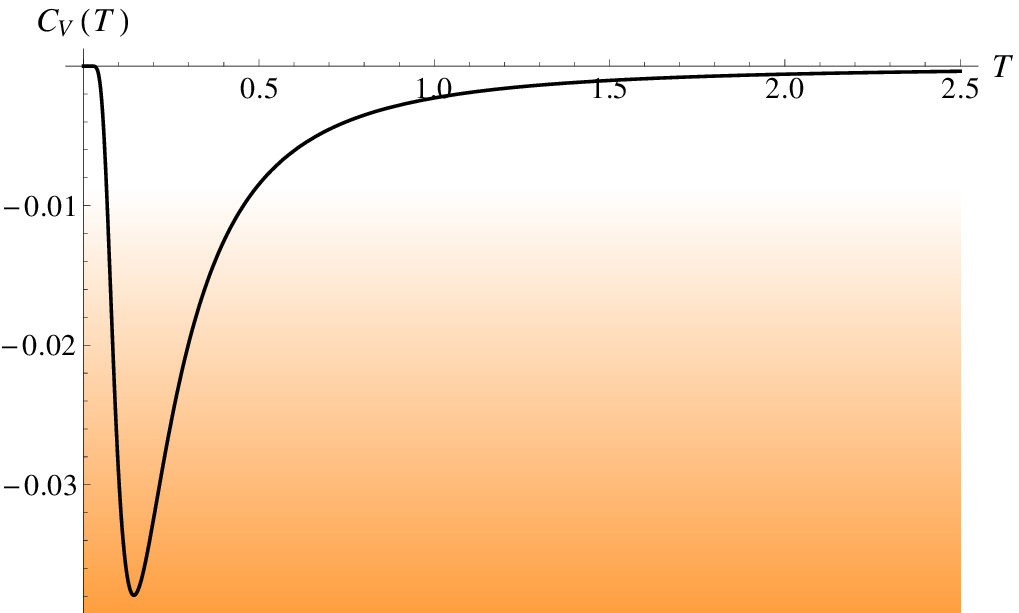}
\caption{The heat capacity (at constant volume) of the quantum vacuum as a function of the temperature, in Planck unities, for $m=0.5 E_P$.}
\end{figure}

\section{Concluding Remarks}

In summary, we have initially presented analytic solutions of the Klein-Gordon equation for an uncharged massive scalar field in the spacetime of a static spherical source (Schwarzschild's spacetime). Then we have drawn the energy eigenvalues of that field in the regime under which $\omega M\rightarrow 0$, so that the particle does not tunnel into the black hole. The consistence of these calculations was shown via non-relativistic limit in which we obtained the gravitational analog of the Bohr's levels valid for the hydrogen atom. It is worth point out that here we are not working with quasi-normal states, and then the complex energies were eliminated by that cutoff condition.

By half-summing the energy eigenvalues and considering the respective degeneracies allowed us to find the vacuum energy associated to the stationary modes of the scalar field, regularizing it by the Riemann's zeta function procedure. The Casimir energy was thus calculated exactly, without taking into account any external boundary, just the spacetime topology itself. Hence we have used the half-sum over the energy eigenvalues instead of local quantities such as the vacuum expected value of the momentum-energy tensor, as it is usual. We have shown that the Casimir energy vanishes when we get the limit $M\rightarrow 0$, but the Casimir tension on the event horizon is non-null.  This fact reveals, in some sense, the nontrivial topology of the spacetime under investigation due to the presence of the singularity at $r=0$. Surprisingly, such a singularity can be identified even when one considers only completely exterior stationary quantum states. 


Finally, we have studied the thermal corrections to the Casimir energy via computation of the Helmholtz free energy in the limit that interest us here, $mM \ll 1$. The thermal Casimir energy also vanishes when the black hole mass tends to zero. However, the thermal tension on the surface of the event horizon once more presents a finite value in this limit, unrevealing again the black hole singularity. However, at high temperatures, this remaining tension on the horizon tends to a constant and minimum absolute value, opposite to the value found in the $T=0$ case. Thus, the total Casimir tension vanishes, which is compatible with the generally accepted catastrophic disappearing of the black hole at its final stages of evaporation, provided the temperature of the thermal bath be that one of the Hawking's radiation, $T_H$, {\it i.e.} $T=T_H\propto1/M$. We also have calculated the Casimir entropy, which is proportional to the horizon area, as the Hawking's (black hole) entropy. That quantity can be in fact interpreted as being the lowest IR contribution to this latter, since it was oobtained from the stationary states of the scalar field. Moreover, we have shown that at low temperatures the Casimir entropy obeys the third law of thermodynamics. At high temperatures, the Casimir entropy also goes to a constant and minimum value. The heat capacity is always negative, showing that, besides the positive values for the Helmholtz free energy, the quantum vacuum is thermodynamically unstable in the considered regime.

As perspectives of the work, we point out the generalization to other quantum fields as well as to stationary and charged spherical gravitational sources, among other solutions. It would be interesting also compare these results with those ones of some effective quantum gravity theories, which predict the existence of a remnant black hole mass that does not evaporate at all \cite{rainbow}. By calling attention to these cases, we hope find a general way of capturing the presence of the singularity via semiclassical gravity, specifically by the global Casimir effect.

\section*{Acknowledgements}

The authors would like to thank Prof. V. B. Bezerra for the fruitful discussions. They also would like to thank Conselho Nacional de Desenvolvimento Cient\'{i}fico e Tecnol\'{o}gico (CNPq) for partial financial support. H.S.V. is funded by the Brazilian research agencies CAPES (PDSE Process No. 88881.133092/2016-01) and CNPq (research Project No. 140612/2014-9).


\end{document}